\documentclass{amsart}
\usepackage[dvips]{color}

\usepackage{lmodern}
\usepackage[T1]{fontenc}
\usepackage{stmaryrd}
\usepackage{amsmath}
\usepackage[foot]{amsaddr}

\definecolor{ddgreen}{rgb}{0.2,0.9,0.0}
\definecolor{turco}{rgb}{0,.5,.5}
\definecolor{reddish}{rgb}{.8,.15,.2}
\definecolor{barney}{rgb}{.6,0.12,0.9}
\definecolor{magic}{cmyk}{0,.7,0,.5}
\definecolor{blueish}{rgb}{.1,.1,.8}
\definecolor{greenish}{rgb}{.2,.4,.2}

\newcommand{\B}{\color{black}}

\begin{document}

\title[Holographic Weyl anomalies]{Simple recipe for holographic Weyl anomaly}
\author{F Bugini$^{\S}$ and D E D\'{\i}az$^{\dag}$}
\address{${\S}$ Departamento de Fisica, Universidad de Concepcion, Casilla 160-C, Concepcion, Chile
${\dag}$ Departamento de Ciencias Fisicas, Universidad
Andres Bello, Av. Republica 252, Santiago, Chile}
\email{${\S}$fbugini@udec.cl, ${\dag}$danilodiaz@unab.cl}

\begin{abstract}
We propose a recipe - arguably the simplest - to compute the holographic type-B Weyl anomaly for general higher-derivative gravity in asymptotically AdS spacetimes. In 5 and 7 dimensions we identify a suitable basis of curvature invariants that allows to read off easily, without any further computation, the Weyl anomaly coefficients of the dual CFT. We tabulate the contributions from quadratic, cubic and quartic purely algebraic curvature invariants and also from terms involving derivatives of the curvature. We provide few examples, where the anomaly coefficients have been obtained by other means, to illustrate the effectiveness of our prescription. 
\end{abstract}


\maketitle
\qquad\\
\section{Introduction}
\qquad\\
The holographic computation of the Weyl (or trace, or conformal) anomaly~\cite{CapperDuff} was one of the early, and certainly most robust,  successes of the AdS/CFT correspondence~\cite{Malda,GKP98,Wit98} in the regime described by the classical Einstein gravity and, correspondingly, leading large-N limit of the strongly coupled CFT. The subtleties of the calculational prescription in identifying the boundary metric $g$ that sourced its dual CFT operator,  namely   the boundary stress-energy tensor, were already examined in the foundational paper~\cite{Wit98} where previous mathematical works in conformal geometry~\cite{FG85,GL91} proved quite helpful. The concrete holographic computations from 3 to 2, 5 to 4 and 7 to 6 dimensions were fully carried out shortly after in the seminal paper~\cite{HS98}. These findings, in turn, sparked the mathematical interest in the volume asymptotics of the Poincar\'e metrics of the Fefferman-Graham construction~\cite{Gra99} (see also~\cite{Albin,CQY}) and it soon paid back~\cite{GZ03,FG02}: the integrated volume anomaly pinpointed the Q-curvature of the boundary manifold, a central construct in conformal geometry~\cite{BO91,Bra93}.

With hindsight, the presence of the four-dimensional Q-curvature in the volume anomaly easily explains the `accidental' equality $a=c$ (see e.g.~\cite{Gubser:1998vd}) between the type-A and type-B conformal anomaly coefficients of the holographic duals of five-dimensional bulk Einstein gravity as follows. In the geometric classification of conformal anomalies~\cite{Deser:1993yx}, the conformal content carried by the Euler density (type-A) is equally captured by the Q-curvature and only the coefficients of the local Weyl-invariant type-B structure get shifted (see e.g.~\cite{GBr05}). In 4D one can trade the Euler density $E_4$ by the Q-curvature ${\mathcal Q}_4$ (type-A) and maintain the Weyl tensor squared $W^2\equiv W_{abcd}W^{abcd}$ which is the obvious independent Weyl-invariant local curvature combination (type-B). This change of basis, modulo a trivial total derivative, amounts to the following apparently trivial rewriting

\begin{eqnarray}
(4\pi)^{2}\,\langle T\rangle&=&-a\,E_4 \,+\,c\,W^2 \\
\nonumber\\
                                            &=&-4 a\,{\mathcal Q}_4 \,+\,(c-a)\,W^2 \nonumber\\\nonumber
\end{eqnarray}
Now, the action for 5D Einstein gravity is just a multiple of the bulk volume and, in consequence, the holographic Weyl anomaly is fully contained in the Q-curvature term. In this case, from the second line of the equation above, one reads off the type-A coefficient $a$ as well as a null coefficient for the (shifted) type-B anomaly $c-a=0$.\\
Likewise, for a 6D CFT one can trade the Euler density $E_6$ by the Q-curvature ${\mathcal Q}_6$~\footnote{We adopt the convention from~\cite{Beccaria:2015ypa}.}

\begin{eqnarray}
(4\pi)^{3}\,\langle T\rangle&=&-a\,E_6\,+\,c_1\,I_1\,+\,c_2\,I_2\,+\,c_3\,I_3 \\
\nonumber\\
\qquad\quad&=&-48a\,{\mathcal Q}_6+(c_1-96a)I_1+(c_2-24a)I_2+(c_3+8a)I_3 \nonumber\\\nonumber
\end{eqnarray}
with the standard basis for type-B conformal anomaly consisting of the three pointwise Weyl invariants $I_1= W_{abcd}W^{eadf}W^{b\;\; c}_{\;ef} \,,\,  I_2=W_{abcd}W^{abef}W_{ef}^{\;\;\;ab}$ and
$I_3=W_{abcd}[\nabla^2\delta^d_{\;e}+4R^d_{\;e} - \frac{6}{5}R \delta^d_{\;e}]W^{abce}+ \mbox{ttd}$. \footnote{Here `ttd' stands for trivial total derivative, in the sense of~\cite{Bastianelli:2000rs}, whose precise form can be found,
for example, in~\cite{BFT00}.}
Again, for bulk 7D Einstein gravity the holographic Weyl anomaly is solely given by the Q-curvature and therefore the three type-B coefficients come out with fixed ratios $c_1:c_2:c_3=12:3:-1$, as follows from the vanishing of the `shifted' type-B coefficients. This combination is rather common in supersymmetric 6D CFTs~\cite{Beccaria:2015ypa} (see also~\cite{Anselmi:1999ut}) and may, eventually, be traced back to the presence of the Q-curvature.

The above observations are also closely  related to a salient feature of the holographic Weyl anomaly of asymptotically Einstein gravity with negative cosmological constant, namely that it vanishes on Ricci-flat boundary metrics. This again can be seen to be inherited from the Q-curvature and, what is more, it can be further shown to be `pure Ricci', i.e. containing only the Ricci tensor and its covariant derivatives. In the explicit results of~\cite{HS98}  this is obvious in 2D and 4D, whereas not immediately apparent in 6D nor in the later reported values in 8D~\cite{Nojiri:2000mk,Gover:2002ay}\footnote{For recent results on the recursion structure of the Q-curvature we refer to~\cite{Juhl11,FG13} and references therein.}. It was then soon realized that in order to have genuine Riemann or, equivalently,  Weyl contributions in the holographic trace anomaly one needs to go beyond volume factorization since, otherwise, pure Ricci terms in the bulk Lagrangian of asymptotically Einstein solutions simply renormalize the radius of the AdS vacuum. Indeed, a subleading correction (order N) to the 4D CFT trace anomaly with $a\neq c$ was obtained from a bulk Riemann-squared term, stemming in turn from type I - heterotic string duality, at the linearized level in the coefficient of this bulk correction to the effective 5D gravitational action~\cite{Blau:1999vz}. Concurrently, the holographic trace anomaly for the most general 5D quadratic bulk Lagrangian was derived in~\cite{Nojiri:1999mh} (see also~\cite{Fukuma:2001uf,Schwimmer:2003eq}) beyond the linearized approximation. In both cases, the coefficients of the type-A and type-B anomaly turned out to be different ($a\neq c$) due to the Riemann-squared term in the bulk Lagrangian.
Incidentally, this very same curvature square correction was shown~\cite{Kats:2007mq,Brigante:2007nu} to lift the universality of the shear-viscosity to entropy-density ratio, $\eta/s=1/4\pi$, holographically obtained for Einstein gravity  duals ~\cite{Policastro:2001yc,Buchel:2003tz}.

Significant progress in the computation of holographic Weyl anomalies was attained in subsequent years. It was mainly prompted by considerations of causality and energy flux positivity constraints and the prospects for new bounds on the viscosity to entropy ratio, the role of supersymmetry and, surprisingly, holographic computations of entanglement entropy.
The perturbative analysis of~\cite{Blau:1999vz} was extended in~\cite{Banerjee:2009fm}  to include six-derivative interactions, whereas the central charges $a$ and $c$ from these 5D curvature-cubed interactions beyond the linearized level were first worked out in~\cite{Myers:2010jv} for the particular combination appearing in quasi-topological gravity~\cite{Myers:2010ru,Oliva:2010eb}. Now, the standard holographic procedure that reconstructs the first few terms of the Fefferman-Graham expansion in terms of the boundary metric by solving analytically the gravitational equation of motion order by order became increasingly difficult to pursue in higher dimensions or in the presence of higher-curvature corrections. Nevertheless, giving up the generality of the boundary metric by restricting to a certain Ansatz for it and resorting to {\sc Mathematica} to solve the equation of motion order by order, it was possible to identify the holographic type-B anomaly coefficients for 7D Gauss-Bonnet gravity~\cite{deBoer:2009pn} and also for 7D cubic Lovelock gravity~\cite{deBoer:2009gx}~\footnote{The type-B coefficients were reported in a simpler form in~\cite{Hung:2011xb} in terms of the renormalized $AdS$ radius. The type-A coefficient reported there followed from the general recipe~\cite{ISTY99}. }. Later on, using the same shortcut route, the type-B anomaly coefficients were computed at the linearized level in the coefficients of the general algebraic quadratic and cubic 7D bulk corrections, as well as of a quartic interaction~\cite{Kulaxizi:2009pz} (see also~\cite{Beccaria:2015ypa}).
By further restricting the boundary metric Ansatz to the product $AdS_2\times S^2$ and assuming a truncated Fefferman-Graham expansion,  a simpler method was found in 5D~\cite{Sen:2012fc} (see also~\cite{Dehghani:2013ldu,Dey:2016pei}) that correctly reproduced the known values of the $CFT_4$ central charges  $a$ and $c$. The extension of this method to 7D~\cite{Sen:2014nfa} required two different boundary product metrics $AdS_2\times S^4$ and $AdS_2\times S^2\times  S^2$ to disentangle the four $CFT_6$ anomaly coefficients and it succeeded, with the aid of {\sc Mathematica} to handle algebraic manipulations, in computing the holographic contribution from terms involving derivatives of the curvature as well.
Finally, let us mention the work~\cite{Miao:2013nfa} in which a different method, based on expansion around a reference background along the lines of~\cite{Blau:1999vz}, was used in order to simplify the derivation of the holographic Weyl anomaly. It contains a rather exhaustive list of holographic contributions from higher-derivative 5D and 7D gravities.

In all of the above-mentioned derivations and shortcut routes to the holographic Weyl anomaly, the challenging part is posed solely by the type-B. It is well known that the holographic type-A Weyl anomaly coefficient is captured by the Lagrangian density evaluated at the AdS solution~\cite{ISTY99}, so that in higher-derivative gravities the only task is to compute the corrected or renormalized AdS radius.  By contrast, the holographic type-B Weyl anomaly lacks this universality and, despite efficient recipes and shortcuts, it seems fair to say that  its derivation is by far more intricate (let alone the possibility to read it off directly from the Lagrangian).

In the present note we show that, contrary to common expectations, the holographic type-B anomaly coefficients can be read off directly from the bulk Lagrangian density. We first allow a mild deviation from the (Euclidean) AdS solution by considering a Poincaré-Einstein bulk metric with an Einstein metric at the conformal boundary, so that the deviations from the volume factorization are due to interactions involving  only contractions of the bulk Weyl tensor and its covariant derivatives which, in turn, can be written explicitly in terms of the boundary Weyl tensor and curvature scalar. It turns out that the restriction to the Einstein condition on the boundary metric does not spoil the independence of the curvature invariants of type-A and type-B trace anomaly, and one can track down their coefficients by looking after the terms logarithmic in the IR-cutoff in the explicit computation of the bulk integrals. We then trade the Euler density by the Q-curvature, since the natural quantity that emerges from the volume anomaly is precisely this pure Ricci combination of curvature invariants, and then identify a suitable basis for the remaining pointwise Weyl invariants that allows to read off their coefficients from the bulk Lagrangian evaluated at the aforementioned Poincaré-Einstein bulk metric.

 The organization of this paper is as follows. We start in Sect.2 with the volume regularization for Einstein bulk metrics with an Einstein metric in the conformal class of its conformal infinity.
 Here we confirm that this restriction correctly reproduces the universal results for type-A and type-B trace anomaly. Section 4 is devoted to the pure Ricci bulk terms that renormalize the radius of AdS.
 In Sect. 4 we compute the holographic contributions of the pure Weyl bulk terms, and again confirm several scattered results in the literature. The restriction to Poincaré-Einstein bulk metric with an Einstein boundary metric is enough to recover the contributions to the `shifted' type-B Weyl anomalies. The suitable basis is identified in Sect. 5 and the simple recipe is unveiled and we tabulate the contributions from the relevant curvature invariants. We illustrate the recipe in Sect. 6 for several celebrated examples. Conclusions are drawn in Sect. 7.
\qquad\\

\section{Poincaré metrics, volume anomaly and Q-curvature}
\qquad\\
Let us start by considering Einstein gravity with negative cosmological constant in $n+1$ dimensions

\begin{equation}
S_{_{EH}}\,=\,\frac{-1}{2\,l^{n-1}_p}\,\int d^{n+1}x\,\sqrt{\hat{g}}\,\{\hat{R}-2\hat{\Lambda}\}
\end{equation}
The negative cosmological constant can be parametrized in terms of a length $L$ that eventually is related to the radius $\tilde{L}$ of the Euclidean AdS solution, i.e. hyperbolic space, $2\hat{\Lambda}=-n(n-1) / L^2$. The solutions, of course, are Einstein bulk metrics with $\tilde{L}=L$ and Ricci tensor
\begin{equation}
\hat{R}ic\,=\,-\frac{n}{\tilde{L}^2}\,\hat{g}
\end{equation}
As per the AdS/CFT correspondence, one examines asymptotically hyperbolic metrics that can be conveniently written in Fefferman-Graham normal form
\begin{equation}
\hat{g}\,=\,\tilde{L}^2\,\frac{dx^2\,+\,g_x}{x^2}
\end{equation}
where the one-parameter family of metrics $g_x$ are to be reconstructed (at least asymptotically) from its boundary value $g$ via the equations of motion. In this way, one meets the Poincaré metrics of the Fefferman-Graham construction and the gravitational action evaluated on these bulk metrics, as a functional of the conformal class $[g]$ of the boundary metric $g$, becomes the partition function of the boundary CFT. The (integrated) holographic Weyl anomaly can then be found from the logarithmically divergent part of the partition function when an IR cutoff $\epsilon$ is introduced.

For Einstein gravity there are two notable exceptions where the Poincaré-Einstein bulk metric can be fully reconstructed from the boundary data $g$.
One is the case of a conformally flat boundary metric~\cite{Skenderis:1999nb} with

\begin{equation}
g_{x}=g\,-\,P\,x^2\,+\,\frac{1}{4}\,P^2\,x^4
\end{equation}
in terms of the Schouten tensor $P$ of the boundary metric $g$.
The second instance, that will be our workhorse for explicit computations, occurs when the conformal class of boundary metrics contains an Einstein metric $g_{_E}$ in which case the dependence on the radial coordinate $x$ factorizes out~\footnote{This was noticed in~\cite{FG12}; however, the authors there remark that this is just a reformulation of a familiar  warped product construction of an Einstein metric in $n+1$ dimensions from an Einstein metric in $n$ dimensions and refer to~\cite{Be}.}
\begin{equation}
g_{x}=\,(1-\lambda\,x^2)^2\,g_{_E}
\end{equation}
where $\lambda$ is just a multiple of the (necessarily constant) Ricci scalar $R$ of $g_{_E}$
\begin{equation}
\lambda\,=\,\frac{R}{4n(n-1)}
\end{equation}
This is the very same expansion as in the conformally flat case, but with the additional input that the Schouten tensor of the Einstein metric $g_{_E}$ is given by $P=\frac{R}{2n(n-1)}g_{_E}$.

Let us now evaluate the Einstein-Hilbert action on a putative~\footnote{In Einstein gravity this is certainly a solution to the equation of motion, but for higher derivative gravities this solution may only exist asymptotically.} Poincaré-Einstein solution with Einstein boundary metric (we abbreviate PE/E). Since the  Lagrangian density is constant, the  bulk volume factorizes out (we decorate $\hat{1}$ to stress this fact)

\begin{equation}
S_{_{EH}}\,=\,\frac{2\hat{\Lambda}-\hat{R}}{2\,l^{n-1}_p}\,\int d^{n+1}x\,\sqrt{\hat{g}}\;\{\hat{1}\}\,=\,
\frac{n}{l^{n-1}_p\,\tilde{L}^2}\,\int d^{n+1}x\,\sqrt{\hat{g}}\;\{\hat{1}\}
\end{equation}
The volume asymptotics can then be explicitly worked out for the PE/E bulk metric
\begin{equation}
vol_{_{PE/E}}\{x > \epsilon\}\,=\,\tilde{L}^{n+1}\int d^nx\, \sqrt{g_{_E}}\,\int_{\epsilon} \frac{dx}{x^{n+1}} \, (1-\lambda\,x^2)^n
\end{equation}
The logarithmic dependence in the cutoff, that we denote ${\mathcal L}_{n+1} \ln \frac{1}{\epsilon}$, originates from the term $x^{-1}\,(-\lambda)^{n/2}$ ${n}\choose{n/2}$ in the expansion of the radial integrand when $n$ is even, and reads
\begin{flalign}
\qquad\qquad{\mathcal L}_{n+1}\,=\, \frac{2\,(-1)^{n/2}\,\tilde{L}^{n+1}}{2^n\,(n/2)!\,(n/2-1)!}\int d^n x\, \sqrt{g_{_E}}\, (n-1)!\,\left[\frac{R}{n(n-1)}\right] ^{ n/2}
\end{flalign}
where the boundary integrand is nothing but the Q-curvature of the boundary Einstein metric $g_{_E}$.

The point of this simple exercise was to ascertain that the restriction to the PE/E bulk metric correctly captures the structure of the volume anomaly and unveils the Q-curvature, in
accordance with the result for a general  $(n+1)-$dimensional Poincar\'e bulk metric as obtained in~\cite{GZ03,FG02}

\begin{flalign}
\qquad\quad
{\mathcal L}_{n+1}\,=\,2\,k_{\frac{n}{2}}\tilde{L}^{n+1}\int d^n x\, \sqrt{g}\,{\mathcal Q}_n~, \qquad\mbox{with} \qquad k_{\frac{n}{2}}=\frac{(-1)^{\frac{n}{2}}}{2^n\frac{n}{2}!(\frac{n}{2}-1)!}
\end{flalign}
If in particular one considers the round metric on the n-sphere at the boundary, then the bulk metric reduces to that of Euclidean $AdS_{n+1}$. Notice that evaluating the action on the $AdS$ vacuum solution still allows access to the type-A anomaly coefficient; this is essentially the content of the universal prescription for type-A holographic Weyl anomaly  first introduced in~\cite{ISTY99} and widely tested and put to work ever since (see e.g.~\cite{Diaz:2007an}-\cite{Beccaria:2015uta}). However, due to the conformal flatness of the round sphere, all type-B terms vanish and their coefficients remain undetermined.
The Q-curvature of the volume anomaly, in turn, contains the combined information on both type-A and type-B trace anomalies since it is a combination of the Euler density (type-A) and the local Weyl invariants (type-B) as follows from the general solution of the Wess-Zumino consistency condition~\cite{Boulanger:2007st} or, in greater generality, from the decomposition of global conformal invariants~\cite{AleI}. Below we examine in more detail the 5D-to-4D and 7D-to-6D cases.
\qquad\\

\subsection{$\bf{AdS_5/CFT_4}$: $\bf{a}$ and $\bf{c}$}
\qquad\\
\qquad\\
Consider the trace anomaly of the 4D CFT

\begin{eqnarray}
(4\pi)^{2}\,\langle T\rangle&=&-a\,E_4 \,+\,c\,W^2 \\
\nonumber\\
\label{4DTA}
                                            &=&-4\,a\,{\mathcal Q}_4 \,+\,(c-a)\,W^2\nonumber\\
\nonumber
\end{eqnarray}
One efficient way to find the two anomaly coefficients $a$ and $c$ is to restrict attention to two particular Einstein spaces (see~\cite{Tseytlin:2013jya} and references therein~\footnote{Alternatively, in the approach of~\cite{Sen:2012fc}, one can also try $AdS_2\times S^2$ with different radii (not Einstein but product of Einstein spaces) and reconstruct the bulk metric with a truncated FG expansion (in this respect, see also~\cite{Gover:2006aw}).}): one conformally (but not Ricci) flat (e.g. the round sphere) and the other Ricci (but not conformally) flat. At the conformally flat metric the Weyl-squared term vanishes and from the non-vanishing Euler density or Q-curvature one obtains $a$; at the Ricci flat metric, on the contrary, the pure Ricci Q-curvature vanishes and one actually determines $c-a$ from the non-vanishing Weyl-squared term.

Our first key observation is that the full information on $a$ and $c$ can be gained by considering the generic Einstein metric $g_{_E}$  :  the Q-curvature reduces to a multiple of the Ricci scalar squared, ${\mathcal Q}_4=R^2/24$, and the Weyl tensor-squared remains unchanged. Therefore, from our previous holographic computation of the volume anomaly for the  5D PE/E bulk metric  that resulted in a coefficient times $R^2$, we readily get $a$ and, since the non-vanishing $W^2$ is absent, it also follows that $c-a=0$.
Plugging back the overall coefficient we read off the holographic Weyl anomaly
\begin{equation}
(4\pi)^2\,\langle T\rangle\,=\,-\frac{4\,\pi^2\,L^3}{l_p^3}\,{\mathcal Q}_4\,=\,-\,N^2\,{\mathcal Q}_4
\end{equation}
In the last equality we translated the Planck length (in units of the AdS radius) to the large-N rank of the gauge group of the CFT$_4$, using the dictionary of the IIB string (SUGRA) $AdS_5\times S^5$ dual to ${\mathcal N}=4$ SU(N) SYM$_4$. Comparing with the expression (\ref{4DTA}), we finally end up with the celebrated result~\cite{HS98}
\begin{equation}
a\,=\,c\,=\,\frac{\pi^2\,L^3}{l_p^3}\,=\,\frac{N^2}{4}
\end{equation}
\qquad\\

\subsection{$\bf{AdS_7/CFT_6}$: $\bf{a, c_1, c_2}$ and $\bf{c_3}$}
\qquad\\
\qquad\\
Consider now the trace anomaly of the 6D CFT

\begin{eqnarray}
(4\pi)^{3}\,\langle T\rangle&=&-a\,E_6\,+\,c_1\,I_1\,+\,c_2\,I_2\,+\,c_3\,I_3 \\
\nonumber\\
\label{6DTA}
\qquad\quad&=&-48a\,{\mathcal Q}_6+(c_1-96a)I_1+(c_2-24a)I_2+(c_3+8a)I_3\nonumber\\\nonumber
\end{eqnarray}
where in the second line we have traded the Euler density by the Q-curvature (modulo a trivial total derivative)
\begin{equation}
E_6\,=\,48\,{\mathcal Q}_6\,+\,96\,I_1\,+\,24\,I_2\,-\,8\,I_3
\end{equation}
so that the Q-curvature captures the type-A content and we have a `shifted' type-B trace anomaly.

In six dimensions things become more involved and restrictions to a conformally flat and a Ricci flat boundary metric do not allow to determine the four anomaly coefficients~\footnote{In the approach of~\cite{Sen:2014nfa}, one can again try products of Einstein spaces and reconstruct the bulk metric with a truncated FG expansion. Two different boundary metrics ($AdS_2\times S^4$ and $AdS_2\times S^2\times S^2$) are required, rendering the procedure more laborious, and also {\sc Mathematica} is used to substitute back in the action to identify the log-divergent term.}, since in that case $E_6=64\,I_1+32\,I_2$ and $I_3=4\,I_1-I_2$.
Even if one considers symmetric Einstein spaces, i.e. with vanishing gradient of the Weyl tensor ($\nabla W=0$) such as $S^6, CP^3, S^2\times S^4,
S^2\times CP^2, S^3\times S^3 \,\mbox{and}\,S^2\times S^2\times S^2$, it is not possible to disentangle the four anomaly coefficients because, as explained in~\cite{Pang:2012rd}, an equality satisfied on symmetric Einstein spaces relates different cubic curvature structures to each other so that there are only three independent cubic curvature structures. More precisely, on symmetric 6D Einstein spaces one has $5\,I_3=32\,I_1 - 8\,I_2$.

In view of the above considerations, we keep a generic 6D Einstein boundary metric $g_{_E}$ so that the Einstein condition reduces the Q-curvature to a multiple of the Ricci scalar cubed,
${\mathcal Q}_6=R^3/225$; the two cubic contractions of the Weyl tensor, that we denote $I_1=W'^3$ and $I_2=W^3$, remain unchanged; whereas the third Weyl invariant reduces to
$I_3=W\nabla^2W - \frac{8}{15} R\,W^2$ modulo the trivial total derivative $\frac{3}{2}\nabla^2W^2$ (see e.g.~\cite{Osborn:2015rna}) that we omit in what follows. The basis of anomaly invariants is then greatly simplified since the Cotton tensor, the Bach tensor and the traceless part of the Ricci tensor all vanish, and yet the four anomaly terms remain independent.
The key observation is then that the outcome of our previous holographic computation of the volume anomaly contains the complete information on both type-A and type-B trace anomaly and allows to determine the four anomaly coefficients. We get the $a$ coefficient in front of the Q-curvature and vanishing  `shifted'  type-B anomaly. Plugging back the overall coefficient, together with the fact that the Q-curvature of the Einstein boundary metric $g_{_E}$ is ${\mathcal Q}_6=R^3/225$, we read off the holographic Weyl anomaly

\begin{equation}
(4\pi)^3\,\langle T\rangle\,=\,\frac{\pi^3\,\tilde{L}^5}{l_p^5}\,{\mathcal Q}_6\,=\, \frac{2\,N^3}{3}\,{\mathcal Q}_6
\end{equation}
In the last equality we again invoked the AdS/CFT dictionary to translate the Planck length (in units of the $AdS_7$ radius) to the large-N rank of the interacting (2,0) superconformal 6D field theory that describes the low energy dynamics of the stack of $M_5$ branes . Finally, comparing with the expression (\ref{6DTA}), we obtain the four anomaly coefficients in agreement with~\cite{BFT00,Beccaria:2015ypa}~\footnote{The apparent mismatch with the original computation in~\cite{HS98} is simply due to the fact that $``I_3 "$ there, actually $M_3$ in the basis of~\cite{Bonora:1985cq}, contains a spurious contribution from the Euler density and it is not classifiable as the third type-B anomaly, as  pointed out  in~\cite{Anselmi:1999ut,Bastianelli:2000rs}.}

\begin{equation}
a\,=\,\frac{c_1}{96}\,=\,\frac{c_2}{24}\,=\,-\frac{c_3}{8}\,=\,-\frac{\pi^3\,L^5}{48\,l_p^5}\,=\,-\frac{N^3}{72}
\end{equation}
\qquad\\

\section{Higher curvature invariants and volume factorization}
\qquad\\
Let us now consider the effect of adding higher curvature terms to the gravitational action.  We start by considering those that are pure Ricci in the bulk, i.e. not involving explicitly the bulk Weyl tensor.  When evaluated on the putative PE/E metric they either vanish, because they contain derivatives of the Ricci tensor  or involve the traceless part of it, or simply reduce to a power of the bulk Ricci scalar that can be written in terms of the renormalized AdS radius. The volume then factors out, so that they will only contribute to the volume anomaly, that is, to the Q-curvature term. The task, then, is simply to determine the renormalized AdS radius and the overall proportionality coefficient, since the ratios between the  type-A and type-B anomaly coefficients remain the same as those from bulk Einstein gravity~\footnote{This is also related to the fact that all these higher curvature invariants can be absorbed by a suitable (local) field redefinition (see e.g.~\cite{Metsaev:1986yb,Banerjee:2009fm,Kats:2007mq,Brigante:2007nu}).}.

We collect all contributions from pure Ricci bulk terms quadratic, cubic and quartic  in the curvature evaluated on the PE/E bulk metric, for which  $\hat{R}ic\,=\,\frac{\hat{R}}{n+1}\,\hat{g}$, and for later convenience we normalize their dimensionless coefficients $u_2, u_3$ and $u_4$ as follows
\begin{flalign}
S[\hat{g}]\,=\,\frac{-1}{2\,l_p^{n-1}}\int d^{n+1}x\,\sqrt{\hat{g}}\,\left\{\frac{n(n-1)}{L^2}\,+\,\hat{R}\,+\,\frac{u_2\,L^2\,\hat{R}^2}{n(n+1)}\,+\,\frac{u_3\,L^4\,\hat{R}^3}{n^2(n+1)^2}\,
+\,\frac{u_4\,L^6\,\hat{R}^4}{n^3(n+1)^3}\right\}
\end{flalign}
where the bulk Ricci scalar contains the renormalized radius $\tilde{L}$ of the AdS vacuum solution
\begin{equation}
\hat{R}\,=\,-\frac{n(n+1)}{\tilde{L}^2}
\end{equation}
as opposed to the length scale $L$ set by the negative cosmological constant
\begin{equation}
\hat{\Lambda}\,=\,-\frac{n(n-1)}{2\,L^2}
\end{equation}
The renormalized AdS radius can then be determined from the trace of the equation of motion. A useful trick~\cite{Hawking:1976ja,Duff:1993wm}, equivalent of course to extremize the action with respect to the AdS radius~\cite{Beccaria:2015ypa}, is to perform a constant scaling of the metric $\hat{g}\rightarrow \mu^2\,\hat{g}$

\begin{flalign}
S[\mu^2\hat{g}]\,=\,\frac{-\mu^{n+1}}{2\,l_p^{n-1}}\int d^{n+1}x\,\sqrt{\hat{g}}\,\left\{\frac{n(n-1)}{L^2}\,+\,\frac{\hat{R}}{\mu^2}\,+\,\frac{u_2\,L^2\,\hat{R}^2}{\mu^4n(n+1)}
\,+\,\frac{u_3\,L^4\,\hat{R}^3}{\mu^6n^2(n+1)^2}\,
+\,\frac{u_4\,L^6\,\hat{R}^4}{\mu^8n^3(n+1)^3}\right\}
\end{flalign}
and then take

\begin{equation}
\frac{\partial S[\mu^2\hat{g}]}{\partial \mu}\mid_{\mu=1}=0
\end{equation}
It is convenient to introduce $f_{\infty}$ to denote the ratio squared $f_{\infty}\,\equiv\,L^2/\tilde{L}^2$, as in~\cite{Hung:2011xb}. The trace of the field equation then demands $f_{\infty}$ to be a positive root of the following polynomial equation
\begin{equation}
0\,=\,1\,-\,f_{\infty}\,+\,\frac{n-3}{n-1}\,u_2\,f_{\infty}^2\,-\,\frac{n-5}{n-1}\,u_3\,f_{\infty}^3\,+\,\frac{n-7}{n-1}\,u_4\,f_{\infty}^4
\end{equation}
Now we substitute back into the action to eliminate $L$ in favor of $\tilde{L}$ and $f_{\infty}$ and factor out the volume

\begin{flalign}
S\,=\,\frac{n}{\tilde{L}^2\,l_p^{n-1}}\,[\,1\,-\,2\,u_2\,f_{\infty}\,+\,3\,u_3\,f^2_{\infty}\,-\,4\,u_4\,f^3_{\infty}\,]
\int d^{n+1}x \sqrt{\hat{g}}\,\{\,\hat{1}\,\}
\end{flalign}
Therefore, the contribution of the pure bulk Ricci terms to the holographic Weyl anomaly  is given by the Q-curvature, as in bulk Einstein gravity.
In 5D we therefore obtain
\begin{flalign}
\setlength\fboxsep{0.3cm}
\setlength\fboxrule{0.5pt}
\boxed{a\,= \, c \,= \,\frac{\pi^2\,\tilde{L}^3}{l_p^3}\bigg[ 1\,- \,2\,u_2\,f_{\infty} \,+\,3\,u_3\,f^2_{\infty}\,- \,4\,u_4\,f^3_{\infty}\bigg]}
\end{flalign}
with
\begin{flalign}
0\,=\,1\,-\,f_{\infty}\,+\,\frac{1}{3}\,u_2\,f_{\infty}^2\,+\,\frac{1}{3}\,u_3\,f_{\infty}^3\,-\,u_4\,f_{\infty}^4
\end{flalign}
Correspondingly, in 7D we obtain
\begin{flalign}
\setlength\fboxsep{0.3cm}
\setlength\fboxrule{0.5pt}
\boxed{
a\,=\,\frac{c_1}{96}\,=\,\frac{c_2}{24}\,=\,-\frac{c_3}{8}\,=\,-\frac{\pi^3\,\tilde{L}^5}{48\,l_p^5}
\bigg[ 1\,- \,2\,u_2\,f_{\infty} \,+\,3\,u_3\,f^2_{\infty}\,- \,4\,u_4\,f^3_{\infty}\bigg]
}
\end{flalign}
with
\begin{flalign}
0\,=\,1\,-\,f_{\infty}\,+\,\frac{3}{5}\,u_2\,f_{\infty}^2\,-\,\frac{1}{5}\,u_3\,f_{\infty}^3\,-\,\frac{1}{5}\,u_4\,f_{\infty}^4
\end{flalign}

In some applications, when the coefficients of the higher curvature terms are parametrically small, it is enough to consider their effect at leading linear order.
It is straightforward to obtain $f_{\infty}$ at linearized level, keeping only terms linear in the $u$'s, by simply substituting the higher powers of $f_{\infty}$ by 1 in the polynomial equation

\begin{flalign}
f_{\infty}\,=\,1\,+\,\frac{n-3}{n-1}\,u_2\,-\,\frac{n-5}{n-1}\,u_3\,+\,\frac{n-7}{n-1}\,u_4
\end{flalign}
The linearized anomaly coefficients are then obtained by substituting this value of $f_{\infty}$ and keeping only $L $ and linear terms in the $u$'s coefficients.
In 5D, this results in

\begin{flalign}
\setlength\fboxsep{0.3cm}
\setlength\fboxrule{0.5pt}
\boxed{
a\,=\,c\,=\,\frac{\pi^2\,L^3}{l_p^3} \left[ 1 - \frac{5}{2}\,(u_2 - u_3 + u_4)\right]
}
\end{flalign}
Correspondingly, in 7D
\vspace{10mm}

\begin{flalign}
\setlength\fboxsep{0.3cm}
\setlength\fboxrule{0.5pt}
\boxed{
a\,=\,\frac{c_1}{96}\,=\,\frac{c_2}{24}\,=\,-\frac{c_3}{8}\,=\,-\frac{\pi^3\,L^5}{48\,l_p^5} \left[ 1\,- \,\frac{7}{2}\,(u_2 - u_3 + u_4) \right]
}
\end{flalign}
The above results are in complete agreement with those of~\cite{Beccaria:2015ypa} and, as expected, the $a$ coefficient also agrees with what predicts the general recipe for type-A~\cite{ISTY99}.
Notwithstanding, we stress that we gained the information on the type-B anomaly as well, all efficiently encoded in the Q-curvature.
\qquad\\

\section{Higher curvature invariants and shifted Type-B}
\qquad\\
Let us now consider curvature invariants in the Lagrangian that cannot be reduced to a factor times the bulk volume when evaluated on the PE/E bulk metric.
We first trade the Riemann tensor by the Weyl tensor, a relation that can be succinctly written in terms of the Kulkarni-Nomizu product
$$\hat{R}iem\,=\,\hat{W}\,+\,\hat{P}\owedge \hat{g}\,=\,\hat{W}\,-\,\frac{1}{2\,\tilde{L}^2}\,\hat{g}\owedge \hat{g}$$
Starting with the most general (local) quadratic curvature  invariant (four-derivative level) the deviation from volume factorization clearly originates from the square of the bulk Weyl tensor.
At the six-derivative level, there are two independent cubic contractions of the bulk Weyl tensor together with two terms involving derivatives of the bulk Weyl tensor.
At quartic level, we restrict the discussion to the purely algebraic contractions.
Let us examine separately the situations in 5D-to-4D and in 7D-to-6D.
\qquad\\

\subsection{5D-to-4D}
\qquad\\
\qquad\\
At the four-derivative level, the nontrivial contribution comes from the following bulk term
\begin{equation}
\Delta S_4\,=\,\int d^{5}x\,\sqrt{\hat{g}}\,\{\,\hat{W}^2\,\}
\end{equation}
This is easy to evaluate on the bulk Poincar\'e-Einstein with Einstein boundary. The warped structure of the bulk metric together with the fact that we have a conformal invariant of weight two,  makes it easy to relate it to the corresponding boundary conformal invariant  $\hat{W}^2\,=\,\tilde{L}^{-4}\,\frac{x^4}{(1-\lambda\,x^2)^4}\,W^2$ so that

\begin{equation}
\Delta S_4\,=\,\tilde{L}\int d^4x\, \sqrt{g_{_E}}\,\int_{\epsilon} \frac{dx}{x} \,W^2
\end{equation}
The crucial observation is then that a bulk $\hat{W}^2$ contributes to the shifted 4D type-B trace anomaly $W^2$
\begin{equation}
(4\pi)^2\,\langle T\rangle\,=\,-16\,\pi^2\,\tilde{L}\,W^2
\end{equation}
Here we already start to grasp the possible simplification that bring in the notions of PE/E metric and Q-curvature; but we still have to examine further curvature terms.

At the six-derivative level, the cubic curvature invariant $\hat{W}^{3'}$, with the same index contraction as the six-dimensional $I_1$, is a conformal invariant of weight three

\begin{equation}
\Delta S_6\,=\,\int d^{5}x\,\sqrt{\hat{g}}\,\{\,\hat{W}^{3'}\,\}
\end{equation}
This is related to the corresponding boundary invariant $I_1=W^{3'}$. On the bulk 5D PE/E metric one has $\hat{W}^{3'}\,=\,\tilde{L}^{-6}\,\frac{x^6}{(1-\lambda\,x^2)^6}\,W^{3'}$, so that
\begin{equation}
\Delta S_6\,=\,\frac{1}{\tilde{L}}\int d^4x\, \sqrt{g_{_E}}\,\int_{\epsilon} dx\frac{x}{(1-\lambda\,x^2)^2} \, {W}^{3'}
\end{equation}
Clearly this term will not produce a logarithmic term and, in consequence, will not contribute to the 4D Weyl anomaly.
The same happens with the second independent cubic curvature invariant $\hat{W}^3$ with the same index contraction as the six-dimensional $I_2$, again a conformal invariant of weight three
\begin{equation}
\Delta S_6\,=\,\int d^{5}x\,\sqrt{\hat{g}}\,\{\,\hat{W}^3\,\}
\end{equation}
This can again be easily related to the corresponding boundary invariant $I_2=W^3$.
On the bulk 5D PE/E metric, $\hat{W}^3\,=\,\tilde{L}^{-6}\,\frac{x^6}{(1-\lambda\,x^2)^6}\,{W}^3$ and therefore

\begin{equation}
\Delta S_6\,=\,\frac{1}{\tilde{L}}\int d^4x\, \sqrt{g_{_E}}\,\int_{\epsilon} dx\frac{x}{(1-\lambda\,x^2)^2} \, {W}^3
\end{equation}
Again, it follows that this term will not produce a logarithmic term and, in consequence, will not contribute to the 4D holographic Weyl anomaly.

By the same token, this will also be the case for any of the seven quartic contractions of the bulk Weyl tensor~\cite{Fulling:1992vm,Boulanger:2004zf}, that we collectively denote by $\hat{W}^4$. They generate no log-term and no contribution to the 4D holographic Weyl anomaly.

The two other possible nontrivial contractions $|\hat{\nabla}\,\hat{W}|^2$ and $\hat{W}\,\hat{\nabla}^2\,\hat{W}$ are connected by a total derivative ($\hat{C_5} \mbox{ in the nomenclature of~\cite{BFT00}}$)
\begin{equation}
|\hat{\nabla}\,\hat{W}|^2 \,+\,\hat{W}\,\hat{\nabla}^2\,\hat{W}\,=\,\frac{1}{2}\hat{\nabla}^2\,\hat{W}^2
\end{equation}
Let us first verify that this total derivative does not contribute to the type-B trace anomaly.
\begin{equation}
\Delta S_6\,=\,\int d^{5}x\,\sqrt{\hat{g}}\,\{\,\hat{\nabla}^2\,\hat{W}^2\,\}
\end{equation}
On the bulk 5D PE/E metric , from its warped structure  together with the conformal properties of the Laplacian and of the Weyl squared term, one has
\begin{equation}
\hat{\nabla}^2\,\hat{W}^2\,=\,\frac{x^6}{\tilde{L}^6\,(1-\lambda\,x^2)^6}\left\{\nabla^2\,+\,\frac{R}{3}\right\}\,W^2
\end{equation}
In 5D no log-term will come out from the radial integration
\begin{equation}
\Delta S_6\,=\,\frac{1}{\tilde{L}}\int d^4x\, \sqrt{g_{_E}}\,\int_{\epsilon} dx\frac{x}{(1-\lambda\,x^2)^2} \, \left\{\nabla^2\,+\,\frac{R}{3}\right\}\,W^2
\end{equation}
and, in consequence, it will not contribute to the 4D holographic trace anomaly.

Lets us now turn to any of the two pieces, say $|\hat{\nabla}\,\hat{W}|^2$ for definiteness (the other one will simply contribute the opposite since the total derivative does not contribute at all)
\begin{equation}
\Delta S_6\,=\,\int d^{5}x\,\sqrt{\hat{g}}\,\{|\hat{\nabla}\,\hat{W}|^2\,\}
\end{equation}
One way to deal with this terms is to use an identity that relates it to the other invariants via a total derivative (corresponding to $\hat{C}_7 \mbox{ in~\cite{BFT00}}$). On the 5D PE/E metric one has
\begin{equation}
\frac{1}{4}|\hat{\nabla}\hat{W}|^2-\frac{2}{\tilde{L}^2}\hat{W}^2-\frac{1}{4}\hat{W}^3+\hat{W}^{3'}=\frac{3}{2}\hat{\nabla}^2\,\hat{W}^2
\end{equation}
Therefore, the contribution is the same as the one from $\frac{8}{\tilde{L}^2}\hat{W}^2$
\begin{equation}
(4\pi)^2\,\langle T\rangle\,=\,\frac {-128\,\pi^2}{\tilde{L}}\, W^2
\end{equation}
We have now exhausted the list of all non-vanishing independent six-derivative invariants, that is, all terms in the A-basis~\footnote{In fact, the B-basis of~\cite{BFT00} is more convenient to evaluate on the PE/E metric. The Ricci tensor is then pure trace and there are only six nontrivial terms, namely, $B_{5}, B_{9}, B_{10}, B_{12}, B_{16}$ and $ B_{17}$.} of~\cite{BFT00}.
We learn therefore that evaluating on the putative PE/E metric provides an easy way to identify the type-A and the type-B holographic Weyl anomaly of the dual $CFT_4$.
\qquad\\

\subsection{7D-to-6D}
\qquad\\
\qquad\\
In 7D the computations are analogous, though a bit more involved.
Let us start with the quadratic Weyl invariant $\hat{W}^2$

\begin{equation}
\Delta S_4\,=\,\int d^{7}x\,\sqrt{\hat{g}}\,\{\,\hat{W}^2\}
\end{equation}
On the 7D PE/E metric $\hat{W}^2\,=\,\tilde{L}^{-4}\,\frac{x^4}{(1-\lambda\,x^2)^4}\,W^2$, so that
\begin{equation}
\Delta S_4\,=\,\tilde{L}^3\int d^6x\, \sqrt{g_{_E}}\,\int_{\epsilon} \frac{dx}{x^3} \, (1-\lambda\,x^2)^2 \,W^2
\end{equation}
The log-term comes exclusively from the following part of the above action
\begin{equation}
-2\tilde{L}^3\int d^6x\, \sqrt{g_{_E}}\,\int_{\epsilon} \frac{dx}{x} \,\lambda\,W^2 \,=\,\frac{-\tilde{L}^3}{60}\int d^6x\, \sqrt{g_{_E}}\,\int_{\epsilon} \frac{dx}{x} \,R\,W^2
\end{equation}
Now one has to express the boundary quantity $R\,W^2$ in terms of the three type-B Weyl invariants  by making use of Bianchi identities and partial integrations .  On the boundary 6D Einstein metric $g_{_E}$, modulo a trivial total derivative, ones has~\footnote{Note that on Ricci flat metrics, the vanishing of the LHS of the above relation imposes a linear dependence between the three Weyl invariants, namely $I_3\,=\,4\,I_1\,-\,I_2$ modulo a trivial total derivative.}
\begin{equation}
\frac{1}{5}\,R\,W^2\,=\,4\,I_1\,-\,I_2\,-\,I_3
\end{equation}
We therefore get the contribution to the trace anomaly
\begin{equation}
(4\pi)^3\,\langle T\rangle\,=\,\frac{16\pi^3\,\tilde{L}^3}{3}(\,4\,I_1\,-\,I_2\,-\,I_3\,)
\end{equation}

Consider now the cubic Weyl invariant $\hat{W}^{3'}$, with the same index contraction as in the six-dimensional $I_1$,

\begin{equation}
\Delta S_6\,=\,\int d^{7}x\,\sqrt{\hat{g}}\,\{\,\hat{W}^{3'}\}
\end{equation}
It is simply related to the corresponding boundary Weyl tensor contraction $W^{3'}$ by the conformal scaling
$\hat{W}^{3'}\,=\,\tilde{L}^{-6}\,\frac{x^6}{(1-\lambda\,x^2)^6}\,W^{3'}$.  It directly produces a log-term

\begin{equation}
\Delta S_6\,=\,\tilde{L}\int d^6x\, \sqrt{g_{_E}}\,\int_{\epsilon} \frac{dx}{x} \,W^{3'}
\end{equation}
and, in consequence, a 6D Weyl anomaly
\begin{equation}
(4\pi)^3\,\langle T\rangle\,=\,-64\pi^3\,\tilde{L}\,I_1
\end{equation}
The same happens with a second independent cubic curvature invariant $\hat{W}^3$, this time the index contraction corresponds to that of the six-dimensional $I_2$ Weyl invariant,
\begin{equation}
(4\pi)^3\,\langle T\rangle\,=\,-64\pi^3\,\tilde{L}\,I_2
\end{equation}

For any of the seven quartic contractions of the bulk Weyl tensor~\cite{Fulling:1992vm,Boulanger:2004zf}, that we generically denote $\hat{W}^4$, we have
\begin{equation}
\Delta S_8\,=\,\int d^{7}x\,\sqrt{\hat{g}}\,\{\,\hat{W}^4\,\}
\end{equation}
When evaluated on the bulk PE/E metric one gets $\hat{W}^4\,=\,\tilde{L}^{-8}\,\frac{x^8}{(1-\lambda\,x^2)^8}\,W^4$, as corresponds to a conformal invariant of weight four,
\begin{equation}
\Delta S_8\,=\,\tilde{L}\int d^6x\, \sqrt{g_{_E}}\,\int_{\epsilon} dx\frac{x}{(1-\lambda\,x^2)^2} \,W^4
\end{equation}
so they will clearly produce no log-term and, therefore, they generate no contribution to the 6D holographic Weyl anomaly.

It remains to examine the two other possible contractions $|\hat{\nabla}\,\hat{W}|^2$ and $\hat{W}\,\hat{\nabla}^2\,\hat{W}$, connected by a total derivative
$|\hat{\nabla}\,\hat{W}|^2 \,+\,\hat{W}\,\hat{\nabla}^2\,\hat{W}\,=\,\frac{1}{2}\hat{\nabla}^2\,\hat{W}^2$~.\\
First, let us examine the fate of the total derivative
\begin{equation}
\Delta S_6\,=\,\int d^{7}x\,\sqrt{\hat{g}}\,\{\,\hat{\nabla}^2\,\hat{W}^2\,\}
\end{equation}
On the bulk 7D PE/E metric , from its warped structure  together with the conformal properties of the Laplacian and of the Weyl squared term, one has
\begin{equation}
\hat{\nabla}^2\,\hat{W}^2\,=\,\frac{x^6}{\tilde{L}^6\,(1-\lambda\,x^2)^6}\left\{\nabla^2\,-\,\frac{8+8\lambda^2 x^4}{x^2}\right\}\,W^2
\end{equation}
In 7D no log-term other than a trivial total derivative $\nabla^2\,W^2$ will come out from the radial integration
\begin{equation}
\Delta S_6\,=\,\frac{1}{\tilde{L}}\int d^4x\, \sqrt{g_{_E}}\,\int_{\epsilon} \frac{dx}{x} \, \left\{\nabla^2\,-\,\frac{8+8\lambda^2 x^4}{x^2} \right\}\,W^2
\end{equation}
and, in consequence, it will not contribute  to the 6D type-B trace anomaly.\\
To deal with any of the individual bulk terms, say $|\hat{\nabla}\,\hat{W}|^2$ for definiteness,
\begin{equation}
\Delta S_6\,=\,\int d^{7}x\,\sqrt{\hat{g}}\,\{|\hat{\nabla}\,\hat{W}|^2\,\}
\end{equation}
we use the identity that relates it to the other invariants via a total derivative ($\hat{C}_7 \mbox{ in~\cite{BFT00}}$) . On the 7D PE/E metric one has
\begin{equation}
\frac{1}{4}|\hat{\nabla}\hat{W}|^2-\frac{3}{\tilde{L}^2}\hat{W}^2-\frac{1}{4}\hat{W}^3+\hat{W}^{3'}=\frac{3}{2}\hat{\nabla}^2\,\hat{W}^2
\end{equation}
Therefore, the contribution to the holographic type-B trace anomaly is the same as the one from $\hat{W}^3-4\hat{W}^{3'}+\frac{12}{\tilde{L}^2}\hat{W}^2$
\begin{equation}
(4\pi)^3\,\langle T\rangle\,=\,64\,\pi^3\tilde{L}\,(\,8\,I_1\,-\,2\,I_2\,-\,I_3\,)
\end{equation}
By now, we have exhausted the list of all non-vanishing independent six-derivative invariants, that is, all terms in the A-basis of~\cite{BFT00}.
Once more we learn that evaluating on the putative PE/E metric provides an easy way to identify the type-A and the type-B holographic Weyl anomaly of the dual $CFT_6$,  but there is yet a drastic simplification to be achieved before we reach our ultimate recipe as we will see in the next section.

\qquad\\

\section{A simple holographic recipe unveiled }
\qquad\\

Our prescription so far consists in evaluating on the putative PE/E bulk metric, with all pure bulk Ricci curvature invariants contributing to the volume anomaly, i.e. to the Q-curvature term in the holographic Weyl anomaly.
We will now add a final refinement and will express the remaining pure Weyl bulk curvature invariants in terms of Weyl invariants of weight two, three and four by making use of Bianchi identities and partial integrations.
In this way, certain combinations of derivative terms may be converted into non-derivative terms and viceversa.
\qquad\\

\subsection{5 to 4 dims}
\qquad\\
\qquad\\
The key observation here is to study a third local Weyl invariant of weight three. We choose,  for example, the one found by  Fefferman and Graham $\Phi$~\cite{FG85} . On the bulk 5D PE/E metric
\begin{equation}
\hat{\Phi}_5\,=\,|\hat{\nabla}\,\hat{W}|^2\,+\,16\,\hat{P}\,\hat{W}^2\,=\,|\hat{\nabla}\,\hat{W}|^2\,-\,\frac{8}{\tilde{L}^2}\,\hat{W}^2
 \end{equation}
$\hat{\Phi}_5$ is on equal footing as the two cubic Weyl contractions. Making use of its conformal covariance under Weyl scaling together with the warped structure of the PE/E metric,  one gets
\begin{equation}
|\hat{\nabla}\,\hat{W}|^2\,-\,\frac{8}{\tilde{L}^2}\,\hat{W}^2\,=\,\frac{x^6}{\tilde{L}^6\,(1-\lambda\,x^2)^6}\left\{|\nabla\,W|^2\,+\,\frac{2}{3}\,R\,W^2\right\}
 \end{equation}
Remarkably, this shows that $\hat{\Phi}_5$ is conformally related to precisely $\Phi_4$ evaluated on the boundary Einstein metric $g_{_E}$
 \begin{equation}
\Phi_4\,=\,|\nabla\,W|^2\,+\,\frac{2}{3}\,R\,W^2
 \end{equation}
The holographic contribution to the Weyl anomaly from this bulk curvature invariant
 \begin{equation}
\Delta S_6\,=\,\int d^{5}x\,\sqrt{\hat{g}}\,\{\,\hat{\Phi}_5\}
\end{equation}
can then be read off from the following explicit result
\begin{equation}
\Delta S_6\,=\,\tilde{L}\int d^4x\, \sqrt{g_{_E}}\,\int_{\epsilon} dx\frac{x}{(1-\lambda\,x^2)^2} \, \left\{\Phi_4\right\}
\end{equation}
It follows then that in 5D no log-term will turn up and we confirm the previously derived result that $|\hat{\nabla}\,\hat{W}|^2$ contributes just as $\frac{8}{\tilde{L}^2}\,\hat{W}^2$~.
However, this information is better encoded in the statement that $\hat{\Phi}_5$ does not contribute to the 4D trace anomaly, just as the  cubic contractions $\hat{W}^{3'}$ and $\hat{W}^3$.\\
With this observation, we finally wind up with the following simple recipe to compute the holographic Weyl anomaly from 5D to 4D :
\begin{itemize}
\item Evaluate on the putative PE/E metric with renormalized AdS radius.\\

\item Write the cosmological constant in terms of the renormalized radius and collect all pure bulk Ricci terms in a constant times the volume (a coefficient times $\hat{1}$).\\

\item Write down the pure bulk Weyl terms (deviations from volume factorization), by means of Bianchi identities and partial integrations, in terms of the Weyl invariant of weight two $\hat{W}^2$, the three Weyl invariants of weight three $\hat{W}^{3'}$, $\hat{W}^3$ and $\hat{\Phi}_5$,  and the Weyl invariants of weight four $\hat{W}^4$.\\

\item Read off the coefficients of the Q-curvature (type-A) and of the Weyl square (shifted type-B) of the holographic Weyl anomaly.
That is, modulo the overall factor in front of the gravitational action, we discover the following simple rule to read off the 4D holographic Weyl anomaly $\mathcal{A}_4$ \B

\begin{flalign}
\setlength\fboxsep{0.3cm}
\setlength\fboxrule{0.5pt}
\boxed{
\mathcal{A}_4\,\left\{ \,\hat{1}\,,\, \hat{W}^2\,,\, \hat{W}^{3'}\,,\,\hat{W}^3,\hat{\Phi}_5\,,\, \hat{W}^4\right\}\,=\,\left\{ \, \frac{1}{16}\,\mathcal{Q}_4\,,\, W^2\,,\, 0\,,\, 0\,,\,0\,,\, 0\right\}
}
\end{flalign}
\end{itemize}

\begin{eqnarray}
(4\pi)^{2}\,\langle T\rangle&=&-4\,a\,{\mathcal Q}_4 \,+\,(c-a)\,W^2\nonumber\\
\end{eqnarray}
It is now simply a bookkeeping exercise to determine the contributions from the general quadratic and cubic curvature invariants and from the ones involving derivatives.
We tabulate the nontrivial ones below.

 \qquad\\
\qquad\\
\[
\begin{array}{|c| c| c|}
\hline
\mbox{Curvature invariant} & \hat{1}/\tilde{L}^4 & \hat{\mathit{W}}^2 \\
\hline \widehat{R}^{\,2}   &400 &- \\
\hline \widehat{R}ic^{\,2}   &80 & - \\
\hline \widehat{R}iem^{\,2}   & 40& 1 \\

\hline
\end{array}
\]
\[
\begin{array}{|r c| c| c| c| c| c|} \hline
& \mbox{Curvature invariant } & \hat{1}/\tilde{L}^6 & \hat{\mathit{W}}^2/\tilde{L}^2 & \hat{W}^{3'}& \hat{W}^{3} &  \hat{\Phi}_5 \\
\hline \widehat{A}_{10}\quad\vline & \widehat{R}^{\,3}   &-8000 & -& -&- &- \\
\hline \widehat{A}_{11}\quad\vline & \widehat{R}\widehat{R}ic^{\,2}   & -1600 & -& -&- &- \\
\hline \widehat{A}_{12}\quad\vline & \widehat{R}\widehat{R}iem^{\,2} & -800 & -20 & -& -& -\\
\hline \widehat{A}_{13}\quad\vline & \widehat{R}ic^{\,3} &-320 & -& -& -& -\\
\hline \widehat{A}_{14}\quad\vline & \widehat{R}iem \, \widehat{R}ic^{\,2} &-320 & -&- &- &- \\
\hline \widehat{A}_{15}\quad\vline & \widehat{R}ic \, \widehat{R}iem^{\,2} &-160 & -4 & -& -& -\\
\hline \widehat{A}_{16}\quad\vline & \widehat{R}iem^{\,3} & -80 & -6 & -&1 &- \\
\hline \widehat{A}_{17}\quad\vline & -\widehat{R}iem'^{\,3} & -60& 3/2& -1 & - &- \\
\hline \widehat{A}_{5}\quad\vline & |\hat{\nabla}\widehat{R}iem|^{2} &- &8 &- &- &1 \\
\hline
\end{array}
\]

\qquad\\

\subsection{7 to 6 dims}
\qquad\\
\qquad\\
The key observation in 7D is to look after a local conformal invariant of weight three, say the Fefferman-Graham invariant $\Phi$. On the bulk PE/E metric we have
\begin{equation}
\hat{\Phi}_7\,=\,|\hat{\nabla}\,\hat{W}|^2\,+\,16\,\hat{P}\,\hat{W}^2\,=\,|\hat{\nabla}\,\hat{W}|^2\,-\,\frac{8}{\tilde{L}^2}\,\hat{W}^2
 \end{equation}
$\hat{\Phi}_7$ is on equal footing as the two cubic Weyl contractions. Making use again of its conformal covariance under Weyl scaling together with the warped structure of the PE/E metric,  one gets
\begin{equation}
|\hat{\nabla}\,\hat{W}|^2\,\,-\,\frac{8}{\tilde{L}^2}\,\hat{W}^2\,=\,\frac{x^6}{\tilde{L}^6\,(1-\lambda\,x^2)^6}\left\{|\hat{\nabla}\,\hat{W}|^2\,+\,\frac{4}{15}\,R\,W^2\right\}
 \end{equation}
and the term into brackets is precisely the 6D invariant $\Phi_6$ evaluated on the boundary Einstein metric $g_{_E}$
 \begin{equation}
\Phi_6\,=\,|\hat{\nabla}\,\hat{W}|^2\,\,+\,\frac{4}{15}\,R\,W^2
 \end{equation}
The holographic contribution to the Weyl anomaly from $\hat{\Phi}_7$ can then be readily obtained from the above result
\begin{equation}
\Delta S_6\,=\,\int d^{7}x\,\sqrt{\hat{g}}\,\{\,\hat{\Phi}_7\}
\end{equation}
so that the logarithmic term can be easily identified 
\begin{equation}
\Delta S_6\,=\,\tilde{L}\int d^6x\, \sqrt{g_{_E}}\,\int_{\epsilon} \frac{dx}{x} \,\Phi_6
\end{equation}
and the corresponding contribution to the type-B Weyl anomaly is given by
\begin{equation}
(4\pi)^3\,\langle T\rangle\,=\,-64\pi^3\,\tilde{L}\,\Phi_6
\end{equation}
With this observation, we finally wind up with the following simple recipe to compute the holographic Weyl anomaly from 7D to 6D :
\begin{itemize}
\item Evaluate on the putative PE/E metric with renormalized AdS radius.\\

\item Write the cosmological constant in terms of the renormalized radius and collect all pure bulk Ricci terms in a constant times the volume (a coefficient times $\hat{1}$).\\

\item Write down the pure bulk Weyl terms (deviations from volume factorization), by means of Bianchi identities and partial integrations, in terms of the three Weyl invariants of weight three $\hat{W}^{3'}$, $\hat{W}^3$ and $\hat{\Phi}_7$,  and the Weyl invariants of weight four $\hat{W}^4$.\\

\item Read off the coefficients of the Q-curvature (type-A) and of the three Weyl invariants (shifted type-B) of the holographic Weyl anomaly.
That is, modulo the overall factor in front of the gravitational action, we discover the following simple rule to read off the 6D holographic Weyl anomaly $\mathcal{A}_6$

\begin{flalign}
\setlength\fboxsep{0.3cm}
\setlength\fboxrule{0.5pt}
\boxed{
\mathcal{A}_6\,\left\{ \,\hat{1}\,,\, \hat{W}^{3'}\,,\,\hat{W}^3,\hat{\Phi}_7\,,\, \hat{W}^4\right\}\,=\,\left\{ \,\frac{-1}{384}\,\mathcal{Q}_6\,,\, I_1\,,\, I_2\,,\, \Phi_6\,\,,\, 0\right\}
}
\end{flalign}
\end{itemize}
We obtain the type-A anomaly coefficient $a$ as well as the three shifted type-B $\tilde{c}_1$, $\tilde{c}_2$ and $\tilde{c}_3$   
\begin{eqnarray}
(4\pi)^{3}\,\langle T\rangle&=&-48a\,{\mathcal Q}_6+\tilde{c}_1I_1+\tilde{c}_2I_2+\tilde{c}_3\Phi_6\\\nonumber\\
\nonumber
&=&-48a\,{\mathcal Q}_6+(c_1+16c_3+32a)I_1+\\
\nonumber\\
\nonumber
&&+(c_2-4c_3-56a)I_2+(3c_3+24a )\Phi_6
\end{eqnarray}
In the second equality we have expressed $\Phi_6$ in terms of the standard basis of Weyl invariants $I_1, I_2$ and $I_3$~\footnote{This relation is exact, modulo a trivial total derivative, for the boundary Einstein metric. In general, this equality involves additional terms containing the Cotton and Bach tensors as well as the traceless part of the Ricci or Schouten tensors.}
\begin{equation}
3\Phi_6=I_3-16I_1+4I_2
\end{equation}

Again, it is now just a matter of bookkeeping to determine the contributions from the general quadratic and cubic curvature invariants and from the ones involving derivatives.
We tabulate the nontrivial ones below.
\qquad\\
\[
\begin{array}{|c| c| c|c|c|}
\hline
\mbox{Curvature invariant } & \hat{1}/\tilde{L}^6 & \;\;\hat{W}^{3'}\;\; & \hat{W}^{3} &  \;\;\hat{\Phi}_7\;\; \\
\hline \widehat{R}^{\,2}/\tilde{L}^2  &1764 &- &- &- \\
\hline \widehat{R}ic^{\,2}/\tilde{L}^2  &252  &-&-&-\\
\hline \widehat{R}iem^{\,2}/\tilde{L}^2  & 84 &1&-1/4&1/4\\
\hline \hat{\mathit{W}}^2/\tilde{L}^2  &-&1&-1/4&1/4\\
\hline
\end{array}
\]

\[
\begin{array}{|r c| c| c| c| c| c|} \hline
& \mbox{Curvature invariant } & \hat{1}/\tilde{L}^6 & \hat{W}^{3'} & \hat{W}^{3} &  \hat{\Phi}_7 \\
\hline \widehat{A}_{10}\quad\vline & \widehat{R}^{\,3}   &-74088 &  -&- &- \\
\hline \widehat{A}_{11}\quad\vline & \widehat{R}\widehat{R}ic^{\,2}  & -10584 & -&- &- \\
\hline \widehat{A}_{12}\quad\vline & \widehat{R}\widehat{R}iem^{\,2} & -3528 & -42 & 21/2& -21/2\\
\hline \widehat{A}_{13}\quad\vline & \widehat{R}ic^{\,3} &-1512 & -& -& -\\
\hline \widehat{A}_{14}\quad\vline & \widehat{R}iem \, \widehat{R}ic^{\,2} &-1512 & -&- &-  \\
\hline \widehat{A}_{15}\quad\vline & \widehat{R}ic \, \widehat{R}iem^{\,2} &-504 & -6 & 3/2& -3/2\\
\hline \widehat{A}_{16}\quad\vline & \widehat{R}iem^{\,3} & -168 & -6 & 5/2&-3/2  \\
\hline \widehat{A}_{17}\quad\vline & -\widehat{R}iem'^{\,3} & -210&1/2& -3/8 & 3/8  \\
\hline \widehat{A}_{5}\quad\vline & |\hat{\nabla}\widehat{R}iem|^{2}   &- &8 &-2 &3  \\
\hline
\end{array}
\]\\


\qquad\\

\section{Comparison to ``experiment''}
\qquad\\
For illustration, we put the recipe to work and report several known instances.  

\qquad\\

\subsection{General quadratic 5D}
\qquad\\
\qquad\\
Historically, this was the first instance where a departure form the $a=c$ feature was obtained.
Let us consider the most general 5D quadratic  Lagrangian
\begin{flalign}
S[\hat{g}]\,=\,\frac{-1}{2\,l_p^3}\int d^5x\,\sqrt{\hat{g}}\,\left\{\hat{R}\,-2\hat{\Lambda}+\,\alpha\,L^2\,\hat{R}^2+\,\beta\,L^2\,\hat{R}ic^{\,2}+\,\gamma\,L^2\,\hat{R}iem^2\right\}
\end{flalign}
Evaluate on the PE/E metric to get
\begin{flalign}
S_{_{PE/E}}\,=\,\frac{-1}{2\,l_p^3}\int d^5x\,\sqrt{\hat{g}}\,\left\{-\frac{20}{\tilde{L}^2}\,+\frac{12}{L^2}\,+\,\frac{400\alpha\,L^2}{\tilde{L}^4}\,+\,\frac{80\beta\,L^2}{\tilde{L}^4}
\,+\,\frac{40\gamma\,L^2}{\tilde{L}^4} \,+\,\gamma\,L^2\,\hat{W}^2\right\}
\end{flalign}
From the trace of the equation of motions it follows the ratio polynomial equation for $f_{\infty}=L^2/\tilde{L}^2$
\begin{flalign}
0\,=\,1\,-\,f_{\infty}\,+\,\frac{20\alpha + 4\beta + 2\gamma}{3}\,f_{\infty}^2
\end{flalign}
Now substitute back in the action to to eliminate $\hat{\Lambda}$ and $L^2$ in favor of $f_{\infty}$ and $\tilde{L}^2$
\begin{flalign}
S_{_{PE/E}}\,=\,\frac{-1}{2\,l_p^3}\int d^5x\,\sqrt{\hat{g}}\,\left\{-8\,\frac{1 - \,(40\alpha + 8\beta + 4\gamma) \,f_{\infty}}{\tilde{L}^2}\,+\,\gamma\,L^2\,\hat{W}^2\right\}
\end{flalign}
Finally, read off the holographic trace anomaly (overall factor $-(4\pi)^2\,\tilde{L}^5$)\\

\begin{eqnarray}
(4\pi)^{2}\,\langle T\rangle&=&-\frac{4\,\pi^2\,\tilde{L}^3}{l_p^3}\,[1\,-\,(40\alpha + 8\beta + 4\gamma) \,f_{\infty}]\,{\mathcal Q}_4 \,+\,\frac{8\,\pi^2\,\tilde{L}^3}{l_p^3}\,\gamma\,f_{\infty}\,W^2\nonumber\\
\end{eqnarray}
 so that
 \begin{flalign}
\setlength\fboxsep{0.3cm}
\setlength\fboxrule{0.5pt}
\boxed{
a\,=\,\frac{\pi^2\,\tilde{L}^3}{l_p^3} \left[1\,-\,(40\alpha + 8\beta + 4\gamma) \,f_{\infty}\right]
}
\end{flalign}
\begin{flalign}
\setlength\fboxsep{0.3cm}
\setlength\fboxrule{0.5pt}
\boxed{
c-a\,=\,\frac{8\pi^2\,\tilde{L}^3}{l_p^3} \,\gamma\,f_{\infty}
}
\end{flalign}
At the linearized level $a$ changes significantly whereas $c-a$ remains pretty much the same since it was already linear in the coefficient $\gamma$
 \begin{flalign}
\setlength\fboxsep{0.3cm}
\setlength\fboxrule{0.5pt}
\boxed{
a\,=\,\frac{\pi^2\,L^3}{l_p^3} \left[1 - 50\alpha - 10\beta - 5\gamma\right]
}
\end{flalign}
\begin{flalign}
\setlength\fboxsep{0.3cm}
\setlength\fboxrule{0.5pt}
\boxed{
c-a\,=\,\frac{8\pi^2\,L^3}{l_p^3} \,\gamma
}
\end{flalign}
In the above derivation one can realize that contrary to the common lore in holographic computations of Weyl anomalies, the (shifted) type-B holographic Weyl anomaly coefficient $c-a$ is determined almost effortlessly.

\qquad\\

\subsubsection{Special case: Gauss-Bonnet}
\qquad\\
\qquad\\
The particular combination of quadratic curvature invariants rendering the Gauss-Bonnet term $\hat{R}^2-4\hat{R}ic^{\,2}+\hat{R}iem^2$ has been extensively studied. This corresponds to
$\alpha=\gamma=\lambda_{_{GB}}/2$ and $\beta=-2\lambda_{_{GB}}$ in the general quadratic Lagrangian. The renormalized AdS radius follows from the positive root of the quadratic equation

\begin{flalign}
0\,=\,1\,-\,f_{\infty}\,+\,\lambda_{_{GB}}\,f_{\infty}^2
\end{flalign}
The anomaly coefficients are then given by
\begin{flalign}
\setlength\fboxsep{0.3cm}
\setlength\fboxrule{0.5pt}
\boxed{
a\,=\,\frac{\pi^2\,\tilde{L}^3}{l_p^3} \left[1\,-\,6\,\lambda_{_{GB}}\,f_{\infty}\right]
}
\end{flalign}
\begin{flalign}
\setlength\fboxsep{0.3cm}
\setlength\fboxrule{0.5pt}
\boxed{
c-a\,=\,\frac{4\pi^2\,\tilde{L}^3}{l_p^3} \,\lambda_{_{GB}}\,f_{\infty}
}
\end{flalign}
At the linearized level $a$ changes significantly whereas $c-a$ remains pretty much the same since it was already linear in the coefficient $\gamma$
\begin{flalign}
\setlength\fboxsep{0.3cm}
\setlength\fboxrule{0.5pt}
\boxed{
a\,=\,\frac{\pi^2\,\tilde{L}^3}{l_p^3} \left[1\,-\,\frac{15}{2}\,\lambda_{_{GB}}\right]
}
\end{flalign}
\begin{flalign}
\setlength\fboxsep{0.3cm}
\setlength\fboxrule{0.5pt}
\boxed{
c-a\,=\,\frac{4\pi^2\,L^3}{l_p^3} \,\lambda_{_{GB}}
}
\end{flalign}

\qquad\\

\subsubsection{Special case: critical gravity 5D}
\qquad\\
\qquad\\
In 5D critical gravity the quadratic curvature correction is given by the Weyl tensor squared with a fine-tuned coefficient
\begin{flalign}
S[\hat{g}]\,=\,\frac{-1}{2\,l_p^3}\int d^5x\,\sqrt{\hat{g}}\,\left\{\hat{R}\,-2\hat{\Lambda}-\frac{L^2}{8}\,\hat{W}^2\right\}
\end{flalign}
The first feature is that since on AdS the Weyl tensor vanishes, there is no correction to the type-A anomaly coefficient and $\tilde{L}=L$ (i.e. $f_{\infty}=1$). The second feature, is that the correction to $c-a$, with $\gamma=-1/8$, is precisely $-a$ so that the type-B anomaly coefficient vanishes 
\begin{flalign}
\setlength\fboxsep{0.3cm}
\setlength\fboxrule{0.5pt}
\boxed{
a\,=\,\frac{\pi^2\,L^3}{l_p^3}\qquad \mbox{and} \qquad c=0
}
\end{flalign}

\qquad\\

\subsection{Most general quadratic and cubic curvature corrections including derivatives in 5D}
\qquad\\
\qquad\\
Let us include now  up to six-derivative invariants in 5D
\begin{equation}
S=-\frac{1}{2 l_p^3}\int d^5x \sqrt{\hat{g}}\left\{\frac{12}{L^2}+\hat{R}+L^2\sum_{i=1}^{3}u_{2,i}\hat{I}_{2,i} +L^4\sum_{j=1}^{8}u_{3,j}\hat{I}_{3,j}+L^4 u_{3,0}\hat{I}_{3,0}\right\}
\end{equation}
where $\hat{I}$'s are the following curvature invariants

\begin{equation}
\begin{array}{c c c c c}\\

\hat{I}_{2,1}=\hat{R}iem^{\,2} & \hat{I}_{2,2}=\hat{R}ic^{\,2} &  \hat{I}_{2,3}=\hat{R}^{\,2} & \hat{I}_{3,1}=\hat{R}iem^{\,3} & \\

\\

\hat{I}_{3,2}=\hat{R}iem^{\,3} +\frac{1}{4}\hat{R}iem^{\,3 '} & \hat{I}_{3,3}=-\hat{R}ic \hat{R}iem^{\,2} & \hat{I}_{3,4}=\hat{R}\hat{R}iem^{\,2} & \hat{I}_{3,5}=\hat{R}iem \hat{R}ic^{\,2}\\

\\

\hat{I}_{3,6}=\hat{R}ic^{\,3} & \hat{I}_{3,7}=\hat{R}\hat{R}ic^{\,2} & \hat{I}_{3,8}=\hat{R}^{\,3} & \hat{I}_{3,0}=|\hat{\nabla}\hat{R}iem|^{\,2}

\\
\end{array}
\end{equation}
It is easy to see that only $\hat{I}_{3,0}$ does not contribute to type-A trace anomaly; whereas  type-B receives no contribution from the pure Ricci $\hat{I}_{2,2}, \hat{I}_{2,3}, \hat{I}_{3,5}, \hat{I}_{3,6}, \hat{I}_{3,7}$ and $\hat{I}_{3,8}$.

\begin{flalign}
0\,=\,1\,-\,f_{\infty}\,+\,\frac{1}{3}\,u_2\,f_{\infty}^2\,+\,\frac{1}{3}\,u_3\,f_{\infty}^3
\end{flalign}
with
\begin{flalign}
20\,u_2\,=\,\frac{1}{10}\,u_{2,1}\,+\,\frac{1}{5}\,u_{2,2}+\,u_{2,3}
\end{flalign}
and
\begin{flalign}
400\,u_3\,=\,\frac{1}{100}\,u_{3,1}\,-\,\frac{1}{200}\,u_{3,2}-\,\frac{1}{50}\,u_{3,3}\,+\,\frac{1}{10}\,u_{3,4}\,+\,\frac{1}{25}\,u_{3,5}\,+\,\frac{1}{25}\,u_{3,6}\,+\,\frac{1}{5}\,u_{3,7}\,+\,\,u_{3,8}
\end{flalign}
We then get for the anomaly coefficients the following

\begin{flalign}
\setlength\fboxsep{0.3cm}
\setlength\fboxrule{0.5pt}
\boxed{
a\,=\,\frac{\pi^2\,\tilde{L}^3}{l_p^3} \left[1\,-\,2\,u_2\,f_{\infty}\,+\,3\,u_3\,f_{\infty}^2\right]
}
\end{flalign}
\begin{flalign}
\setlength\fboxsep{0.3cm}
\setlength\fboxrule{0.5pt}
\boxed{
c-a\,=\,\frac{8\pi^2\,\tilde{L}^3}{l_p^3} \,f_{\infty} \left[v_2\,-\,v_3\,f_{\infty}\right]
}
\end{flalign}
with
\begin{flalign}
v_2\,=\,u_{2,1}
\end{flalign}
and
\begin{flalign}
v_3\,=\,6\,u_{3,1}\,+\,3\,u_{3,2}-\,4\,u_{3,3}\,+\,20\,u_{3,4}\,-\,8\,u_{3,0}
\end{flalign}
\\
At the linearized level, we have to expand the $f_{\infty}$'s  in the brackets and $f_{\infty}^{-3/2}$ from the $\tilde{L}^3$ in front

\begin{flalign}
f_{\infty}\,=\,1\,+\,\frac{1}{3}\,u_2\,+\,\frac{1}{3}\,u_3
\end{flalign}

\begin{flalign}
\setlength\fboxsep{0.3cm}
\setlength\fboxrule{0.5pt}
\boxed{
a\,=\,\frac{\pi^2\,L^3}{l_p^3} \left[1\,-\,\frac{5}{2}\,u_2\,+\,\frac{5}{2}\,u_3\right]
}
\end{flalign}
\begin{flalign}
\setlength\fboxsep{0.3cm}
\setlength\fboxrule{0.5pt}
\boxed{
c-a\,=\,\frac{8\pi^2\,L^3}{l_p^3} \left[v_2\,-\,v_3\right]
}
\end{flalign}

All the above results are in perfect agreement with what has been reported in the literature~\cite{Kulaxizi:2009pz,Miao:2013nfa,Beccaria:2015ypa} .
\qquad\\

\subsubsection{Special case: quasi-topological gravity}
\qquad\\
\qquad\\
Let us close 5D by considering the particular the case of quasi-topological gravity 
\begin{flalign}
S[\hat{g}]\,=\,\frac{-1}{2\,l_p^3}\int d^5x\,\sqrt{\hat{g}}\,\left\{\frac{12}{L^2}\,+\,\hat{R}\,+\,\frac{\lambda\,L^2}{2}\,\hat{\mathcal{X}}_4\,+\,\frac{7\,\mu\,L^4}{8}\,\hat{\mathcal{Z}}'_5\right\}
\end{flalign}
where $\hat{\mathcal{X}}_4$ is the Gauss-Bonnet term 
\begin{equation}
\hat{\mathcal{X}}_4\,=\,\hat{R}^2\,-4\,\hat{R}ic^{\,2}+\,\hat{R}iem^2
\end{equation}
and $\hat{\mathcal{Z}}'_5$ is the purely algebraic cubic invariant
\begin{equation}
\hat{\mathcal{Z}}'_5\,=\,\hat{R}iem^3\,+\,\frac{1}{14}\left( 21\hat{R}\hat{R}iem^2\, -\,120\hat{R}ic\hat{R}iem^2\, +\,144\hat{R}iem\hat{R}ic^2 \, +\,128\hat{R}ic^3 
\, -\,108\hat{R}\hat{R}ic^2\, +\,11\hat{R}ic^3\right)
\end{equation}
Evaluating on the PE/E bulk metric, with the input from the table where the A-basis for cubic curvature invariants is evaluated,  one gets
\begin{flalign}
0\,=\,1\,-\,f_{\infty}\,+\,\lambda\,f_{\infty}^2\,+\,\mu\,f_{\infty}^3
\end{flalign}
Now we substitute back in the action to eliminate $L$ in favor of $\tilde{L}$ and $f_{\infty}$ and obtain
\begin{flalign}
S[\hat{g}]\,=\,\frac{-1}{2\,l_p^3}\int d^5x\,\sqrt{\hat{g}}\,\left\{\frac{-8\,+\,48\,\lambda\,f_{\infty}\,-72\,\mu\,f^2_{\infty}}{\tilde{L}^2}\,\hat{1}\,+\,\frac{\lambda\,-3\,\mu\,f_{\infty}}{2}\,f_{\infty}\,\tilde{L}^4\,\hat{W}^2
\,+\,\frac{7\,\mu\,f^2_{\infty}}{8}\,\tilde{L}^6\,\hat{W}^3\right\}
\end{flalign}
Our recipe now instructs us how to read off the holographic Weyl anomaly 
\begin{eqnarray}
(4\pi)^{2}\,\langle T\rangle&=&-\frac{4\,\pi^2\,\tilde{L}^3}{l_p^3}\,[1\,-\,6\,\lambda\,f_{\infty}\,+9\,\mu\,f^2_{\infty}]\,{\mathcal Q}_4 \,+\,\frac{4\,\pi^2\,\tilde{L}^3}{l_p^3}\,[\lambda\,-3\,\mu\,f_{\infty}]\,f_{\infty}\,W^2\nonumber\\
\end{eqnarray}
 so that
 \begin{flalign}
\setlength\fboxsep{0.3cm}
\setlength\fboxrule{0.5pt}
\boxed{
a\,=\,\frac{\pi^2\,\tilde{L}^3}{l_p^3} \left[1\,-\,6\,\lambda\,f_{\infty}\,+9\,\mu\,f^2_{\infty}\right]
}
\end{flalign}
\begin{flalign}
\setlength\fboxsep{0.3cm}
\setlength\fboxrule{0.5pt}
\boxed{
c-a\,=\,\frac{4\pi^2\,\tilde{L}^3}{l_p^3} \,[\lambda\,-3\,\mu\,f_{\infty}]\,f_{\infty}
}
\end{flalign}
in conformity with~\cite{Myers:2010jv}.
\qquad\\

\subsection{Most general quadratic and cubic curvature corrections including derivatives in 7D}
\qquad\\
\qquad\\
Let us turn now to 7D and include up to six-derivative invariants 
\begin{equation}
S=-\frac{1}{2 l_p^5}\int d^7x \sqrt{\hat{g}}\left\{\frac{30}{L^2}+\hat{R}+L^2\sum_{i=1}^{3}u_{2,i}\hat{I}_{2,i} +L^4\sum_{j=1}^{8}u_{3,j}\hat{I}_{3,j}+L^4 u_{3,0}\hat{I}_{3,0}\right\}
\end{equation}
As before, only $\hat{I}_{3,0}$ does not contribute to type-A trace anomaly; whereas  type-B receives no contribution from the pure Ricci $\hat{I}_{2,2}, \hat{I}_{2,3}, \hat{I}_{3,5}, \hat{I}_{3,6}, \hat{I}_{3,7}$ and $\hat{I}_{3,8}$.

The polynomial equation for the renormalized AdS radius is given by
\begin{flalign}
0\,=\,1\,-\,f_{\infty}\,+\,\frac{3}{5}\,u_2\,f_{\infty}^2\,-\,\frac{1}{5}\,u_3\,f_{\infty}^3
\end{flalign}
with
\begin{flalign}
42\,u_2\,=\,\frac{1}{21}\,u_{2,1}\,+\,\frac{1}{7}\,u_{2,2}+\,u_{2,3}
\end{flalign}
and
\begin{flalign}
1764\,u_3\,=\,\frac{1}{441}\,u_{3,1}\,-\,\frac{1}{441}\,u_{3,2}-\,\frac{1}{147}\,u_{3,3}\,+\,\frac{1}{21}\,u_{3,4}\,+\,\frac{1}{49}\,u_{3,5}\,+\,\frac{1}{49}\,u_{3,6}\,+\,\frac{1}{7}\,u_{3,7}\,+\,u_{3,8}
\end{flalign}
After substituting back in the action to eliminate $L$ in favor of $\tilde{L}$ and $f_{\infty}$, we follow our recipe to read off the type-A anomaly coefficient
\begin{flalign}
\setlength\fboxsep{0.3cm}
\setlength\fboxrule{0.5pt}
\boxed{
a\,=\,-\frac{\pi^3\,\tilde{L}^5}{48 l_p^5} \left[1\,-\,2\,u_2\,f_{\infty}\,+\,3\,u_3\,f_{\infty}^2\right]
}
\end{flalign}
and the shifted type-B anomaly coefficients
\begin{flalign}
\setlength\fboxsep{0.3cm}
\setlength\fboxrule{0.5pt}
\boxed{
\tilde{c}_1\,=\,\frac{32\pi^3\,\tilde{L}^5}{l_p^5} \,f_{\infty} \left[u_{2,1}\,-\,6\,u_{3,1}\,f_{\infty}-\,2\,u_{3,2}\,f_{\infty}+\,6\,u_{3,3}\,f_{\infty}-\,42\,u_{3,4}\,f_{\infty}+\,8\,u_{3,0}\,f_{\infty}\right]
}
\end{flalign}

\begin{flalign}
\setlength\fboxsep{0.3cm}
\setlength\fboxrule{0.5pt}
\boxed{
\tilde{c}_2\,=\,-\frac{8\pi^3\,\tilde{L}^5}{l_p^5} \,f_{\infty} \left[u_{2,1}\,-\,10\,u_{3,1}\,f_{\infty}-\,4\,u_{3,2}\,f_{\infty}+\,6\,u_{3,3}\,f_{\infty}-\,42\,u_{3,4}\,f_{\infty}+\,8\,u_{3,0}\,f_{\infty}\right]
}
\end{flalign}

\begin{flalign}
\setlength\fboxsep{0.3cm}
\setlength\fboxrule{0.5pt}
\boxed{
\tilde{c}_3\,=\,\frac{8\pi^3\,\tilde{L}^5}{l_p^5} \,f_{\infty} \left[u_{2,1}\,-\,6\,u_{3,1}\,f_{\infty}-\,3\,u_{3,2}\,f_{\infty}+\,6\,u_{3,3}\,f_{\infty}-\,42\,u_{3,4}\,f_{\infty}+\,12\,u_{3,0}\,f_{\infty}\right]
}
\end{flalign}
At the linearized level, we have to expand the $f_{\infty}$'s  in the brackets and $f_{\infty}^{-5/2}$ from the $\tilde{L}^5$ in front
\begin{flalign}
f_{\infty}\,=\,1\,+\,\frac{3}{5}\,u_2\,-\,\frac{1}{5}\,u_3
\end{flalign}

\begin{flalign}
\setlength\fboxsep{0.3cm}
\setlength\fboxrule{0.5pt}
\boxed{
a\,=\,-\frac{\pi^3\,L^5}{48 l_p^5} \left[1\,-\,147\,u_2\,+\,6174\,u_3\right]
}
\end{flalign}
\begin{flalign}
\setlength\fboxsep{0.3cm}
\setlength\fboxrule{0.5pt}
\boxed{
\tilde{c}_1\,=\,\frac{32\pi^3\,L^5}{l_p^5} \left[u_{2,1}\,-\,6\,u_{3,1}\,-\,2\,u_{3,2}\,+\,6\,u_{3,3}\,-\,42\,u_{3,4}\,+\,8\,u_{3,0}\right]
}
\end{flalign}

\begin{flalign}
\setlength\fboxsep{0.3cm}
\setlength\fboxrule{0.5pt}
\boxed{
\tilde{c}_2\,=\,-\frac{8\pi^3\,L^5}{l_p^5} \left[u_{2,1}\,-\,10\,u_{3,1}\,-\,4\,u_{3,2}\,+\,6\,u_{3,3}\,-\,42\,u_{3,4}\,+\,8\,u_{3,0}\right]
}
\end{flalign}

\begin{flalign}
\setlength\fboxsep{0.3cm}
\setlength\fboxrule{0.5pt}
\boxed{
\tilde{c}_3\,=\,\frac{8\pi^3\,L^5}{l_p^5} \left[u_{2,1}\,-\,6\,u_{3,1}\,-\,3\,u_{3,2}\,+\,6\,u_{3,3}\,-\,42\,u_{3,4}\,+\,12\,u_{3,0}\right]
}
\end{flalign}
We again confirm that all the above results are in perfect agreement with what has been reported in the literature~\cite{Kulaxizi:2009pz,Miao:2013nfa,Beccaria:2015ypa} .
\qquad\\

\subsubsection{Special case: Lovelock 7D}
\qquad\\
\qquad\\
For concreteness, let us illustrate how the recipe works in the particular the case of 7D (Lanczos-)Lovelock gravity 
\begin{flalign}
S[\hat{g}]\,=\,\frac{-1}{2\,l_p^5}\int d^7x\,\sqrt{\hat{g}}\,\left\{\frac{30}{L^2}\,+\,\hat{R}\,+\,\frac{\lambda\,L^2}{12}\,\hat{\mathcal{X}}_4\,-\,\frac{\mu\,L^4}{24}\,\hat{\mathcal{X}}_6\right\}
\end{flalign}
where $\hat{\mathcal{X}}_4$ is the Gauss-Bonnet term 
\begin{equation}
\hat{R}^2\,-4\,\hat{R}ic^{\,2}+\,\hat{R}iem^2
\end{equation}
and $\hat{\mathcal{X}}_6$ is the purely algebraic cubic invariant that in six dimensions corresponds to the Euler density

\begin{equation}
4\,\hat{R}iem^3 + 8\hat{R}iem^{3 '} + 3\hat{R}\hat{R}iem^2 - 24\hat{R}ic\hat{R}iem^2 + 24\hat{R}iem\hat{R}ic^2  + 16\hat{R}ic^3 - 12\hat{R}\hat{R}ic^2 + \hat{R}ic^3
\end{equation}
\\
Evaluating on the 7D PE/E bulk metric, with the input from the table where the A-basis for cubic curvature invariants is evaluated,  one gets
\begin{flalign}
0\,=\,1\,-\,f_{\infty}\,+\,\lambda\,f_{\infty}^2\,+\,\mu\,f_{\infty}^3
\end{flalign}
Now we substitute back in the action to eliminate $L$ in favor of $\tilde{L}$ and $f_{\infty}$ and obtain for the integrand

\begin{flalign}
\frac{-12\,+\,40\,\lambda\,f_{\infty}\,+180\,\mu\,f^2_{\infty}}{\tilde{L}^2}\,\hat{1}\,+\,\frac{\lambda\,+\,5\,\mu\,f_{\infty}}{12}\,f_{\infty}\,\tilde{L}^4\,\hat{W}^{3 '}\,+
\end{flalign}
\begin{flalign}
\nonumber
-\,\frac{\lambda\,+\,17\,\mu\,f_{\infty}}{48}\,f_{\infty}\,\tilde{L}^4\,\hat{W}^3\,+ \,\frac{\lambda\,+\,9\,\mu\,f_{\infty}}{48}\,f_{\infty}\,\tilde{L}^4\,\hat{\Phi}_7
\end{flalign}
\\
Our recipe now instructs us how to read off the holographic Weyl anomaly. We obtain the type-A anomaly coefficient 
\begin{flalign}
\setlength\fboxsep{0.3cm}
\setlength\fboxrule{0.5pt}
\boxed{
a\,=\,-\frac{\pi^3\,\tilde{L}^5}{48 l_p^5} \left[1\,-\,\frac{10}{3}\,\lambda\,f_{\infty}\,-\,15\,\mu\,f_{\infty}^2\right]
}
\end{flalign}
and the shifted type-B anomaly coefficients
\begin{flalign}
\setlength\fboxsep{0.3cm}
\setlength\fboxrule{0.5pt}
\boxed{
\tilde{c}_1\,=\,\frac{8 \pi^3\,\tilde{L}^5}{3\,l_p^5} \,f_{\infty} \left[\lambda\,+\,5\,\mu\,f_{\infty}\right]
}
\end{flalign}

\begin{flalign}
\setlength\fboxsep{0.3cm}
\setlength\fboxrule{0.5pt}
\boxed{
\tilde{c}_2\,=\,-\frac{2 \pi^3\,\tilde{L}^5}{3\,l_p^5} \,f_{\infty} \left[\lambda\,+\,17\,\mu\,f_{\infty}\right]}
\end{flalign}

\begin{flalign}
\setlength\fboxsep{0.3cm}
\setlength\fboxrule{0.5pt}
\boxed{
\tilde{c}_3\,=\,\frac{2 \pi^3\,\tilde{L}^5}{3\,l_p^5} \,f_{\infty}  \left[\lambda\,+\,9\,\mu\,f_{\infty}\right]}
\end{flalign}
in conformity with~\cite{Hung:2011xb}.
\qquad\\

\subsection{Quartic quasi-topological Gravity 5D}
\qquad\\
\qquad\\

Our last example contains purely algebraic quartic curvature invariants and corresponds to 5D quartic quasi-topological gravity. We refer to~\cite{Dehghani:2013ldu} for details. 
To be brief, we just tabulate the contributions from the quartic terms evaluated on the 5D PE/E metric 

\[
\begin{array}{|r c| c| c| c| c| c|} \hline
& \mbox{Curvature invariant } & \hat{1}/\tilde{L}^8 & \hat{\mathit{W}}^2/\tilde{L}^4 & \hat{W}^{3'}/\tilde{L}^2 & \hat{W}^{3}/\tilde{L}^2 &  \hat{\Phi}_5/\tilde{L}^2 \\
\hline & \widehat{R}iem^{\,4}   &160 & 24 & -8 &- &- \\
\hline  & \widehat{R}iem^{\,2}\widehat{R}ic^{\,2}   & 3200 & 80& -&- &- \\
\hline & \widehat{R}\widehat{R}ic^{\,3} & 6400 & - & -& -& -\\
\hline & (\widehat{R}iem^{\,2})^{\,2} &1600 & 80& -& -& -\\
\hline & \widehat{R}ic^{\,4} &1280 & -&- &- &- \\
\hline & \widehat{R}\widehat{R}iem \widehat{R}ic \widehat{R}ic &6400 & - & -& -& -\\
\hline  & \widehat{R}iem \widehat{R}ic\widehat{R}ic\widehat{R}ic & 1280 & - & -&- &- \\
\hline & \widehat{R}iem\widehat{R}iem\widehat{R}ic\widehat{R}ic & 320& 16& - & - &- \\
\hline  & \widehat{R}^4 &160000 &- &- &- &- \\
\hline  & \widehat{R}iem^{\,2}\widehat{R}^2 &16000 &400 &- &- &- \\
\hline  & \widehat{R}ic^{\,2}\widehat{R}^2 &32000 &- &- &- &- \\
\hline  & \widehat{R}iem\widehat{R}iem\widehat{R}iem\widehat{R}ic &320 &24 &- &- &- \\
\hline  & \widehat{R}iem^{'\,4}&280 &6 &- &4 &- \\
\hline
\end{array}
\]
\qquad\\
With the input from the table above, one gets
\begin{flalign}
0\,=\,1\,-\,f_{\infty}\,+\,\lambda\,f_{\infty}^2\,+\,\mu\,f_{\infty}^3\,+\,\nu\,f_{\infty}^4
\end{flalign}
Now one substitutes back in the action to eliminate $L$ in favor of $\tilde{L}$ and $f_{\infty}$. Then it only remains to track down the coefficients of the volume $\hat{1}$ and of $\hat{W}^2$ to read off 
 \begin{flalign}
\setlength\fboxsep{0.3cm}
\setlength\fboxrule{0.5pt}
\boxed{
a\,=\,\frac{\pi^2\,\tilde{L}^3}{l_p^3} \left[1\,-\,6\,\lambda\,f_{\infty}\,+9\,\mu\,f^2_{\infty}\,+4\,\mu\,f^3_{\infty}\right]
}
\end{flalign}
\begin{flalign}
\setlength\fboxsep{0.3cm}
\setlength\fboxrule{0.5pt}
\boxed{
c-a\,=\,\frac{4\pi^2\,\tilde{L}^3}{l_p^3} \,[\lambda\,-3\,\mu\,f_{\infty}\,-2\,\mu\,f^2_{\infty}]\,f_{\infty}
}
\end{flalign}
in conformity with~\cite{Dehghani:2013ldu}~\footnote{
In more generality, a full list of quartic curvature invariants can be found in~\cite{Amsterdamski:1989bt,Fulling:1992vm}. 
There is no obstruction to compute the contributions from purely algebraic curvature invariants. However, to properly take into account terms with derivatives of the Riemann tensor we would need the basis of Weyl invariants of weight four~\cite{Boulanger:2004zf}. This will be discussed somewhere else.}.
\qquad\\

\section{Conclusion}
\qquad\\
We have found an easy way to compute the holographic Weyl anomaly in theories with higher curvature invariants. It has long been known that the holographic type-A trace Weyl anomaly can be obtained from the bulk action evaluated on the AdS solution. We have now allowed a small departure by considering a putative Poincaré-Einstein bulk metric with an Einstein metric at the conformal boundary; the action greatly simplifies and by splitting volume terms coming from pure Ricci bulk invariants from pure Weyl bulk deviations one is able to read off the holographic type-B Weyl anomaly. The key point is to express the pure Weyl bulk terms in a basis of Weyl invariant bulk terms. Having done that, the natural basis for the CFT trace anomaly is the one where one trades the Euler density by the pure Ricci Q-curvature. The bulk volume naturally `descends' to the Q-curvature and the bulk Weyl invariants `descend' to the corresponding Weyl invariants of the type-B Weyl anomaly.       
\qquad\\

\section*{Acknowledgement}
\qquad\\
We are grateful to R.Aros, N.Boulanger, R.Olea, J.Oliva and S.Theisen for valuable discussions.
We also acknowledge useful email correspondence with A.Chang, K.Sen and A.Sinha.
F.B. is grateful to K.Peeters for assistance with CADABRA code.
D.E.D. thanks M.Selva for assistance with LaTeX.
The work of F.B. was partially funded by grant CONICYT-PCHA/Doctorado Nacional/2014-21140283.
D.E.D. acknowledges support from project UNAB DI 1271-16/R.
\qquad\\


\providecommand{\href}[2]{#2}\begingroup\raggedright\endgroup

\end{document}